\documentclass{article}
\usepackage{fullpage}
\usepackage{amsfonts,amssymb}
\usepackage{amsthm}
\usepackage{tikz}
\usetikzlibrary{calc}
\usetikzlibrary{arrows}
\usepackage{tikz-3dplot}
\usepackage{color} 
\usepackage[fleqn]{amsmath}
\usepackage{amsthm}
\usepackage{amssymb}
\usepackage{amsfonts}
\usepackage{bbm}
\usepackage{epsfig}
\usepackage{times}
\usepackage{hyperref}
\usepackage{bm}
\usepackage{float} 
\usepackage{appendix} 
\usepackage{lscape}
\usepackage{cite}
\usepackage{mathrsfs}
\usepackage{appendix}
\usepackage{setspace}
\usepackage{mathtools}
\usepackage{mdframed}
\usepackage{authblk}
\theoremstyle{definition}

\newtheorem{definition}{Definition}[section]

\usepackage{tikz}

\begin{document}

\title{Does Science need Intersubjectivity? The Problem of Confirmation in Orthodox  Interpretations of Quantum Mechanics}

 \author{Emily Adlam  \thanks{The Rotman Institute of Philosophy, 1151 Richmond Street, London N6A5B7 \texttt{eadlam90@gmail.com} }}

\date{\today}

\maketitle

\vspace{5mm}

What must an interpretation of quantum mechanics do in order to be considered viable? We suggest that one vitally important criterion is the following: \emph{Any successful interpretation of quantum mechanics must explain how our empirical evidence allows us to come to know about quantum mechanics.} That is, an interpretation of quantum mechanics must be able to tell a sensible story about how empirical confirmation works in the context of quantum-mechanical experiments, otherwise the whole project will be self-defeating: we cannot rationally believe an interpretation of a theory which tells us we have no good reason to believe that the theory itself is right, because our only reason for believing the interpretation is our belief that the theory is right! 

 The `probability problem' and the related problem of confirmation has been much-discussed in the context of the Everett interpretation\cite{AdlamEverett, Greaves2006-GREOTE,kent2009world,Sebens2016-SEBSUA, Wallace}, but it is evident that other interpretations which also postulate major changes to our usual ideas about the relationship between observers and reality will  be vulnerable to similar objections. In this article, we will focus on the class of interpretations sometimes known as  `orthodox' interpretations\cite{cuffaro2021open} or alternatively `Copenhagenish' interpretations\cite{Leifertalk}. These are interpretations which posit nothing but `\emph{unitary dynamics to characterize the dynamical evolution of a state vector}' and which `\emph{deny that we should think of a system as having an observer-independent state.}'\cite{cuffaro2021open} Orthodox interpretations differ from the Everett interpretation because they tell us that measurements have unique outcomes, but those outcomes are relativized to an observer. Examples of orthodox interpretations include   Copenhagen interpretations\cite{Bohr1987-BOHTPW,Heisenberg1958-HEIPAP,pauli1994writings}, neo-Copenhagen interpretations\cite{Zeilinger1999-ZEIAFP,Zeilinger2002,brukner2015quantum,articlebanana,demopoulos2012generalized,Janas2021-JANUQR,2004neoc}, QBism\cite{QBismintro}, pragmatic interpretations\cite{Healey2012-HEAQTA} and some versions of relational quantum mechanics or RQM\cite{1996cr} (specifically, those which postulate that unitary quantum mechanics is `complete').\footnote{ A companion paper (ref \cite{pittphilsci20379}) sets out a proposal to update relational quantum mechanics to address the problems raised in this article. Moreover, in a sense this paper is just making explicit ideas which were implicit in RQM all along, so one might contend that even earlier versions of RQM are not really vulnerable to our criticisms either. However, this depends on the way in which they are formulated - in particular, we will see that the assumption that `unitary quantum mechanics is complete' is the source of much of the problem, and there are earlier versions of RQM which explicitly postulate that unitary quantum mechanics is complete, so our criticisms do apply to those formulations.} Certain forms of the `It from Bit' hypothesis may also fall into this category, insofar as they conceive of `information'  in an epistemic, agent-relative way rather than as a part of physical reality\cite{Wheeler1989-WHEIPQ,Kastrup2019-KASAIA-3}\footnote{For reasons of space we will assume readers are familiar with the interpretations we discuss in this paper, and thus we will not provide any detailed explication of them  beyond our characterisation of general features of operational interpretations, but readers who wish to know more are invited to consult the texts we have referenced here.}. We emphasize that the class of interpretations we are concerned with in this paper does not include instrumentalist approaches in which one simply chooses not to   ask questions about the reality underlying quantum mechanics - our criticisms pertain specifically to orthodox interpretations which claim to be revealing a deep truth about reality, e.g. that it is intrinsically perspectival or that there is no observer-independent description of it. 
 
It has been observed  that interpretations of this kind  challenge the standard scientific doctrine of   \emph{intersubjectivity} about measurement outcomes and other macroscopic events\cite{2021quintet,doi:10.1093/bjps/axp017,Ruyant2018CanWM,vanFraassen2010-VANRQM,Mucino2022-MUCARQ}, i.e. they imply that measurements and other macroscopic events will not typically have the same outcome for all macroscopic observers. Proponents of orthodox interpretations have historically taken a rather cavalier attitude toward this fact - for after all, every interpretation of quantum mechanics has some features which appear strange to our classical intuition, so it might seem that orthodox interpretations are no worse off than other approaches in this regard. However, what this discussion  overlooks is the fact that intersubjectivity plays a vitally important role in the process of empirical confirmation, and thus the failure of intersubjectivity in orthodox interpretations means that these approaches have serious difficulties in fulfilling the crucial criterion of showing how our empirical evidence could allow us to know about quantum mechanics.  This is not just a matter of learning to accept something which clashes with our classical intuitions: if no reasonable account of empirical confirmation can be given within these approaches, it would be irrational to believe in the picture of the world that they present to us, regardless of how appealing or unappealing we might otherwise find it. 

 We will begin in section \ref{orthodox} by demonstrating how intersubjectivity fails within orthodox interpretations of quantum mechanics, and then we will  present two related argument aiming to show that these failures of intersubjectivity undermine the process of empirical confirmation. First, in section \ref{OM}, we will argue that  observers in the kind of reality postulated by the orthodox interpretations would be unable to escape their own perspective in order to learn anything about the perspectives of other observers, and thus they would be unable to empirically confirm quantum mechanics unless the theory is regarded as nothing more than a description of the contents of a single perspective. Second, in section \ref{past}, we will argue that since there is no natural criterion of identity for `observers' in orthodox interpretations, we cannot assume that observers  in this kind of reality persist over time, and thus these observers would not even be able to confirm quantum mechanics as a description of their own personal experiences. We will conclude in section \ref{RQM} by discussing what could be done to save the orthodox interpretations, concluding that it cannot be rational to believe  these sorts of interpretations unless they are supplemented with   some observer-independent structure which underwrites intersubjective agreement between different observers or at least between successive `observers' associated with what we would normally take to be the same person; and we suggest how this can be achieved in the case of RQM. 
  
To be clear,  our concern in this article is not about whether or not   it's possible to find empirical evidence that favours an orthodox interpretation over other interpretations of quantum mechanics - roughly speaking, we accept that if  epistemic concerns of the kind raised in this article can be cleared up, then all interpretations of quantum mechanics will make the same predictions at least within the domain of non-relativistic quantum mechanics. Our concern is rather that if we believe one of the orthodox interpretations, then quantum mechanics itself will lose its status as empirically confirmed, and then there can be no possible reason to believe in the orthodox interpretation in question, or indeed to believe in any   interpretation of quantum mechanics at all. That is, we are worried about the  very \emph{possibility} of empirical confirmation within the kind of universe postulated by the orthodox interpretations.

Note that in  this text we will largely address the orthodox interpretations together, aiming our arguments at key shared elements which we take to be responsible for the main epistemic problems. In some cases we will make digressions to discuss the example of a specific orthodox approach, but we will not work through the arguments separately for every different orthodox interpretation -    there are simply too many of them for this to be practical. We consider that this level of detail is  adequate   to demonstrate that all of these interpretations have some questions to address, but we  acknowledge that there may be relevant distinctions which entail that our arguments do not land in quite the same way in the context of some specific orthodox interpretations, so we would invite proponents of orthodox interpretations to consider these matters in more detail.

  \section{Intersubjectivity in Orthodox Interpretations \label{orthodox}}

We define the class of interpretations we are concerned with - `orthodox interpretations' - by their adherence to the following principles (we note that Pienaar uses a very similar set of principles to define `Copenhagenish' interpretations\cite{2021pienaaragain}):

\begin{itemize} 

\item Relative to each observer, every measurement has a unique outcome, and  the probabilities for these unique outcomes are given by the Born rule - i.e. there are no Everett-style branches.

\item  Unitary quantum mechanics is universal and complete - i.e. nothing needs to be added to the unitary formalism, there is no `collapse of the wavefunction,' there are no hidden variables, and  there is no explicit mechanism for selecting and actualising a single measurement outcome either relative to an observer or absolutely. 

\item Quantum mechanics does not describe an observer-independent external reality - i.e. there is no observer-independent fact of the matter about the true quantum state of a given system at a given time.

\end{itemize}

All orthodox interpretations have  different ways of characterising what they take quantum mechanics to  be about, given that it is not about an observer-independent reality, but the general picture is that it describes  facts relative to an observer, agency, consequences, effects, experiences or similar - the key point is that the content of quantum mechanics is understood to be relativized to an observer or agent, and is often regarded as being epistemic in some way. For example, `\emph{A QBist takes quantum mechanics to be a personal mode of thought – a very powerful tool that any agent can use to organize her own experience ...  quantum mechanics itself does not deal directly with the objective world; it deals with the experiences of that objective world that belong to whatever particular agent is making use of the quantum theory.}'\cite{2020qbism} Similarly,  Brukner writes of his neo-Copenhagen interpretation that `\emph{measurement records   ... can have meaning only relative to the observers; there are no “facts of the world per se,”}'\cite{brukner2015quantum}, Zeilinger emphasizes that `\emph{it is information about possible measurement results that is represented in the quantum states,}'\cite{Zeilinger2002} and Timpson summarizes the view of Zeilinger as the statement that `\emph{a physical system literally is nothing more than an agglomeration of actual and possible sense impressions arising from observations}'\cite{Timpson2003-TIMOAS}. Note that in this article we will use the general term `perspectives' to refer indiscriminately to  all of these similar ontologies -   we understand the `perspective' of an observer to contain facts relative to the observer, experiences, consequences, measurement outcomes, effects, or whatever it is that a given orthodox interpretation takes the subject of quantum mechanics to be.

We note that while some orthodox interpretations (e.g. the neo-Copenhagen approaches of Brukner\cite{brukner2015quantum} and Zeilinger\cite{Zeilinger1999-ZEIAFP} and some versions of RQM\cite{Ruyant2018CanWM}) deny that there is any observer-independent reality, or `view from nowhere,' at all,  others (e.g. QBism\cite{QBismintro}) maintain that there is an external reality but  deny that quantum mechanics functions as a third-person description of it. Orthodox approaches of the latter kind necessarily insist that the nature of external reality is  unspeakable and ineffable, in the sense that we cannot have information about it, obtain evidence about it or meaningfully form hypotheses about any of its features other than those revealed to us directly in our own experience in the form of measurement outcomes. For example,  Timpson suggests we should understand QBism as motivated by the view that `\emph{once one seeks to go beyond a certain level of detail, the world simply does not admit of any straightforward description or capturing of by theory}'\cite{Timpson2008-TIMQBA}\footnote{We note in passing that if this is the motivation for QBism, then one might think that QBism should be adopted only as a last resort when  all attempts to describe an underlying reality have conclusively failed, and we are not convinced that this level of despair is justified by the current state of the field.}. Thus QBists cannot use their putative observer-independent reality to insist that, contra the discussion above, there is some  observer-independent fact  about which outcome of a measurement has really occurred, because if that were the case then there would be a fact of the matter about which quantum state assignment were right in this scenario, and QBism insists firmly that there is never any fact of the matter about which quantum state assignment is right.  Fuchs emphasizes this point in his discussion of Wigner's friend: `\emph{Who has the right state of information? The conundrums simply get too heavy if one tries to hold to an agent-independent notion of correctness for otherwise personalistic quantum states.}'\cite{2010arXiv1003.5209F}  

Admittedly QBists do hold out the hope that  `\emph{when the subjective Bayesian conception of quantum states and related structures is adopted it will be possible to `see through' the quantum formalism to the real ontological lessons it is trying to teach us}'\cite{Timpson2008-TIMQBA} but with these kinds of claims they are walking a fine line - they must maintain that   as a matter of principle this underlying ontology  can be grasped only dimly and we will never be able to describe it or say anything concrete about it,  since otherwise QBism would just amount to a hidden-variable approach with the stipulation that we don't yet have a good model of the hidden variables. If there are QBists out there who believe that the underlying reality is not ineffable and we can meaningfully form hypotheses about it and/or come up with a theory describing it, our arguments would not apply to their position - but we would ask QBists of that genre why they are content to stop at QBism rather than trying to understand the nature of the underlying reality.

Thus although certain orthodox interpretations admit the existence of an observer-independent reality, if they are to remain orthodox interpretations it is crucial   that they should forbid us from articulating any concrete hypotheses about the nature of that observer-independent reality,  and therefore the putative existence of the observer-independent reality doesn't significantly change the epistemic status of these approaches as compared to  orthodox interpretations which deny observer-independent reality altogether.

We emphasize that our arguments in this paper are not intended to apply to approaches which merely hold that quantum mechanics has \emph{some} epistemic or relational aspects. We agree with Jaynes that `\emph{[O]ur present [quantum mechanical] formalism  ... is a peculiar mixture describing in part realities of Nature, in part incomplete human information about Nature - all scrambled up by Heisenberg and Bohr into an omelette that nobody has seen how to unscramble,}'\cite{Jaynesomelette} and therefore we agree that at least some parts of the formalism should be regarded as  `epistemic' and/or  `relative to an observer.' We also recognise that one might reasonably  believe that some parts of quantum mechanics are epistemic   while being uninterested in the nature of the underlying non-epistemic     reality, so we have no problem with the orthodox interpretations if they are regarded as merely `good enough for now' or `good enough for most current applications' - e.g. we would not object to a deflationary  form of QBism which suggests that we should interpret quantum mechanics as  a normative principle for making predictions based on what we currently know, with the proviso that one day it will hopefully be supplanted by a description of what is really going on underneath. We object only to views which insist that there is no observer-independent reality at all, or that observer-independent reality is  in principle unknowable and  ineffable; and in particular we object to the corollary that science should not even attempt to describe an observer-independent reality which could be associated with quantum mechanics. For it is our contention that the rationality of scientific confirmation depends implicitly on the assumption that there exists an observer-independent reality which regulates and coordinates different perspectives, and therefore one cannot consistently conclude on the basis of a scientific theory that there is no observer-independent reality or that nothing whatsoever  can be known about it\footnote{We note that in some of the examples we have listed above as orthodox interpretations, it is not entirely obvious whether their proponents intend to say a) there is no observer-independent reality, b) observer-independent reality is ineffable or c)  observer-independent reality  isn't ineffable, but quantum mechanics isn't a description of observer-independent reality,  so we can apply it sensibly even if we don't have an understanding of the nature of the underlying reality. The  arguments of this article apply to these views only if they are interpreted as saying either a) or b) - we have no complaint about c).}.

\subsection{Failures of  Intersubjectivity \label{wide}}

The point of relativizing the content of quantum mechanics to an observer, as proposed by the orthodox interpretations, is that  this makes it possible to sidestep  some of the  problems of standard quantum mechanics by insisting that  `\emph{in quantum mechanics different observers may give different accounts of the same sequence of events.}'\cite{1996cr} This is supposed to help answer the questions raised by the `Wigner's friend' scenario and other similar cases. For example, suppose Bob knows that his friend Alice is performing a measurement on  a system S which probes the variable $V$ of $S$.  When Alice performs the measurement, she witnesses some measurement outcome $M_A$ and thus learns the value of $V$. But since Bob describes the whole interaction unitarily, from his point of view the interaction has only caused Alice and S to become entangled; it has not selected any one measurement result. Orthodox interpretations resolve the tension between these two descriptions by insisting that Alice's measurement has a definite outcome relative to  Alice but not relative to Bob. 

We reinforce that in order for this strategy to work, the orthodox interpretations must maintain that the observer-dependence of the content of quantum mechanics applies even to macroscopic events like measurement outcomes, as this is `\emph{the very injunction that keeps the potentially conflicting statements of (Bob) and (Alice) in check'}\cite{2010arXiv1003.5209F}. Indeed, since orthodox interpretations tell us that quantum mechanics is universal, they cannot possibly insist that measurement outcomes are observer-independent while microscopic quantum events are observer-dependent - that would be to postulate an in-principle split between the microscopic and the macroscopic, which is exactly what orthodox interpretations seek to avoid. Thus such interpretations must accept that different observers can assign different states to the apparatus after a measurement, and thus observers can disagree about which measurement outcome has occurred, or about whether any definite outcome has occurred at all - and crucially,  the orthodox interpretations insist that there is  no fact of the matter about which observer  is right.  So in orthodox interpretations, the fact that some observer has seen some measurement  outcome does not entail that this outcome has occurred in any objective, observer-independent sense, or that it will be reflected in anyone else's perspective. And by and large, proponents of orthodox interpretations seem to accept this consequence: Healey observes that `\emph{QBists and others have come to question and even deny the principle that a well-conducted quantum measurement has a definite, objective, physical outcome}'\cite{2018qtatloo} and proponents of orthodox interpretations including Brukner\cite{2018angtfoif} and Healey\cite{2018qtatloo} have presented several no-go theorems leading to the conclusion that `\emph{the universal applicability of unitary quantum theory is inconsistent with the assumption that a well-conducted measurement always has a definite physical outcome}'\cite{2018qtatloo}\footnote{As detailed in refs \cite{2018qtatloo,Baumann2019-BAUCOH-3} there are reasons one might object to the  arguments of both Brukner and Healey. We mention them here not because we endorse them but simply as evidence that proponents of orthodox interpretations typically accept that measurements do not always have definite outcomes.}.

Having acknowledged this, we have some further questions to resolve about the scenario of Alice and Bob. Suppose that Bob measures S in the $V$ basis, and hence he obtains a measurement outcome $M_B^S$ which he will interpret as providing information about the result of Alice's measurement of $V$ on S.  Suppose that Bob also `measures' Alice herself (e.g. perhaps he simply asks her about her measurement result) and obtains a measurement outcome $M_B^A$ for the value of some pointer variable which is supposed to be a record of her measurement result. So in this scenario we have three different measurement outcomes $M_A, M_B^S, M_B^A$ all supposedly providing information about the value of the same variable.  What should an orthodox interpretation say about the relationships between these three measurement results? 

We present two different postulates which could be employed to answer this question.  We have drawn  these postulates from discussions in the literature on RQM (\textbf{Shared facts} is the version proposed by ref \cite{2021quintet}, while \textbf{Internally consistent descriptions} is the version which appears in \cite{pittphilsci19664}) but it is clear that other orthodox interpretations will also face a similar dilemma about how to describe this scenario.    

\begin{definition}  \textbf{Shared facts (SF):} If Bob measures  Alice to ‘check the reading’ of a pointer variable, the value he finds is necessarily equal to the value that Alice recorded in her earlier measurement of S (i.e. $M_A = M_B^A$)

\end{definition} 

\begin{definition} 

\textbf{Internally consistent descriptions (ICD):} If Bob measures Alice then also measures S, the two values found are in agreement (i.e. $M_B^A = M_B^S$)

\end{definition}

 We note that \textbf{ICD} actually follows directly from the formalism of unitary quantum mechanics. For when Alice measures S, from Bob's point of view she becomes entangled with S in a state of the form $\sum_i c_i   |i \rangle_S | i \rangle_A$ where $ \{ | i \rangle_S \}$ is the set of possible values for the variable $V$, and $| q \rangle_A$ represents the state of Alice in which she has seen the outcome $V = q$. Clearly then if Bob measures S and gets outcome $| q \rangle_S$, when he measures Alice he must get outcome $| q \rangle_A$,  and vice versa, so $M_B^S$ and $M_B^A$ must always agree - in particular, if Bob `measures' Alice by asking her what her measurement outcome was, necessarily Bob will experience Alice as reporting a measurement result which matches the result that Bob himself obtained in his measurement on S. Thus since orthodox interpretations make no changes to the standard framework of unitary quantum mechanics, they must all predict that \textbf{ICD} is true, i.e. they all entail that $M_B^S = M_B^A$. (We note that the case of QBism is a little more complicated since QBism tells us that the probabilities in quantum mechanics are not objective - we will return to this in a moment).

We now have a  further question about whether \textbf{SF} is also true. If \textbf{ICD} is true but \textbf{SF} is not true, then it could be the case that $M_B^S = M_B^A$ but $M_A \neq M_B^S $ and hence $M_A \neq M_B^A$: in that case, whatever Alice tells Bob about the result of her experiment will be perceived by Bob as being consistent with Bob's other observations, but what Bob perceives Alice as saying may be different from what Alice experienced herself as observing and what she perceives herself as saying. 

And in fact, a short argument shows that \textbf{SF} cannot be reliably true within any orthodox interpretation: 
 
 \begin{enumerate} 
 
 \item Orthodox interpretations tell us that   there are no observer-independent facts about events which can be described quantum-mechanically.
 
 \item  Orthodox interpretations insist that quantum mechanics is universal, so all macroscopic events can be described quantum-mechanically; thus orthodox interpretations tell us that there are no observer-independent facts about  \emph{macroscopic} events.
 
 \item Since there are no observer-independent facts about macroscopic events in the context of an orthodox interpretation, then in order to say that \textbf{SF} is reliably true in such an interpretation, we will need the interpretation to explicitly  provide some structure which coordinates measurement outcomes for different observers  in the way specified by \textbf{SF}.
  
 \item Unitary quantum mechanics does not have the resources to ensure that the measurement outcomes for different observers will reliably match in the way specified by \textbf{SF}, since it does not even have an explicit mechanism to select and actualise a single measurement outcome relative to any one observer. 
  
 \item  Orthodox interpretations do not add any structures or mechanisms to unitary quantum mechanics, so they don't have the resources to ensure that the measurement outcomes for different observers will reliably match in the way specified by \textbf{SF}.
 
 \item So orthodox interpretations cannot tell us that \textbf{SF} is reliably true.
 
 \end{enumerate} 
 
Let us consider points three and four in more detail. First, evidently  in an interpretation which allows the existence of observer-independent facts, we can have an explicit  mechanism similar to a wavefunction collapse which selects and actualises a unique observer-independent  measurement outcome, and that outcome can play a coordinating role which  ensures that  if observers   use their perceptual equipment and communicate properly, their outcomes will match in the way specified by \textbf{SF}. But orthodox interpretations do not postulate any observer-independent facts, so  they must appeal to some other kind of structure if they are to uphold \textbf{SF}. Since the orthodox interpretations insist that unitary quantum mechanics is universal and complete, the only structures they endorse are those inherent in the unitary dynamics of quantum mechanics. But the unitary dynamics do not  even provide any mechanism to select and actualise a single measurement outcome relative to one observer, so they certainly cannot provide a mechanism to select and actualise a single measurement result in a global way that holds for all observers, so  there is no structure available within the orthodox interpretations which could ensure that \textbf{SF} is upheld.  In an interpretation like the Everett interpretation we can get around this problem by relativizing \textbf{SF} to a branch of the wavefunction and showing that  the measurement results of all observers  within a single branch  will always match. However,  since orthodox interpretations insist that measurement have unique outcomes, this route is not open to them either. Thus it would seem in an orthodox interpretation there is simply nothing that could possibly play the role of making sure that    $M_B^S$ matches $M_A$,  and for the same reasons there is nothing  that could possibly play the role of making sure that  $M_B^A$ matches $M_A$; thus \textbf{SF} cannot be reliably true within an orthodox interpretation. This is also the conclusion reached by Van Fraassen, who notes that `\emph{we have no basis or law  (within unitary quantum mechanics) on which to connect the outcomes of measurements by different observers, no matter how intimately they may be related.}'\cite{vanFraassen2010-VANRQM}   Similarly, Mu\~{c}ino et al note that standard quantum mechanics as used in RQM `\emph{explicitly lacks such a single unified description, i.e., it does not contain a mathematical object encoding all the different perspectives.}' \cite{Mucino2022-MUCARQ}

We recognise that it may be responded by some proponents of orthodox interpretations that  questions about the relation between $M_A$ and $M_B^A$  are illegitimate - i.e. it is meaningless to ask if Alice and Bob's perspectives agree, because `\emph{“facts” can only exist relative to the observer'}\cite{brukner2015quantum}. It will presumably be argued that within the context of an orthodox interpretation, the closest we could manage to such a comparison would be to have some third observer measure Alice and Bob separately and compare the results, and of course due to \textbf{ICD} these results will always be found to agree\footnote{Again, we acknowledge that QBism does not necessarily entail that ICD holds so QBism does not necessarily tell us that these results will always be found to agree.}, so from this point of view there can never be any genuine disagreement between Alice and Bob. However, the results of a measurement by a third observer tells us nothing about the relationship between the subjective experiences of Alice and Bob, and subjective experiences cannot be disregarded,  because it is our subjective experiences which form the basis for the empirical confirmation of a scientific theory like quantum mechanics.   In fact we fully concur that  questions about the relation between relationships between the subjective experiences of Alice and Bob are probably meaningless if the world is the way that the orthodox interpretations say it is - but that is precisely the problem! As we will shortly argue, it is necessary to be able to pose and answer such questions if we are to carry out empirical confirmation for a scientific theory like quantum mechanics, so if the orthodox interpretations say that such questions can't meaningfully be posed, then the orthodox interpretations necessarily tell us that quantum mechanics is not empirically confirmed. Our complaint is not simply that the orthodox interpretations do not have answers to questions that we would like answered - our complaint is that if these questions really can't be posed then most forms of scientific confirmation will be severely undermined.

\paragraph{ICD} 

We note that \textbf{ICD} has been presented by some proponents of orthodox interpretations as a solution to the problem of intersubjectivity: so for example Laudisa and Rovelli write, `\emph{this means that a third system measuring X and $F_X$ will certainly find consistent values. That is: the perspectives of F and W agree on this regard, and this can be checked in a physical interaction.}'\cite{sep-qm-relational} But in fact, \textbf{ICD} is really the source of the problem! For  if quantum mechanics is indeed universal, as the orthodox interpretations insist, it follows that when one  observer asks another observer what outcome they obtained in some measurement, that is just another kind of quantum measurement - it is a physical interaction in which one observer `measures' the other observer by asking questions about their measurement results. And thus  since we have already granted that in orthodox interpretations there is no guarantee that different observers will agree on the outcome of a quantum measurement,  it follows that they may also disagree about what is communicated in this sort of conversation. Furthermore, not only is it possible that they will disagree,  \textbf{ICD} tells us that they \emph{must} disagree any time they compare notes about the result of a measurement where their respective perspectives say different things about the outcome of the measurement. For example, we have seen that in the case of Alice and Bob, unitary quantum mechanics cannot provide a guarantee that $M_B^S$ matches $M_A$; and since \textbf{ICD} entails that Bob's two outcomes match, it follows that if Alice's result is different from Bob's, then  when Bob asks Alice about the result of her measurement (i.e. when he performs the measurement whose result we have labelled $M_B^A$),  necessarily Alice and Bob will disagree about the very words that Alice says during their conversation! And presumably, since the orthodox interpretations insist that everything is always relativized to an observer, the same is true no matter how Alice and Bob attempt to communicate with one another: there is no way for them to get outside their own perspectives to achieve intersubjective agreement, and \textbf{ICD} will always prevent them from telling one another about the ways in which their relative descriptions disagree. So Alice and Bob seem to be permanently stuck in incommensurate realities - they can never actually compare their descriptions of reality because all possible means of communication are also part of the relative description. Thus orthodox interpretations are saddled with ubiquitous failures of intersubjectivity that can't be rectified by any amount of communication and comparison. Pienaar poetically describes this scenario as  an ontology of `\emph{island universes.}'\cite{2021quintet}

The situation is slightly different in QBism (and in any other orthodox interpretation which tells us that the probabilities associated with quantum mechanics are purely subjective). Like the other orthodox interpretations QBism says that quantum mechanics is complete and universal, and Pienaar emphasizes that this means `\emph{it can be used upon any part of reality (including other agents),}' so it remains true that in the context of QBism   Bob must necessarily describe Alice after her measurement using the entangled state  $\sum_i c_i | i \rangle_S | i \rangle_A$. It then seems natural to conclude that Bob's measurements must match and thus \textbf{ICD} must hold in QBism, meaning that agents in a QBist universe will likewise be prevented from communicating their experiences to each other. However,  because the probabilities in QBism are purely subjective probabilities, it does not follow from Bob's assignation of the state $\sum_i c_i | i \rangle_S | i \rangle_A$ that the results of Bob's measurements must actually match, because Bob's assignation of this state only describes his own subjective beliefs and does not place any constraints on what will actually happen.  So although it is the case that QBism, like all orthodox interpretations, must \emph{predict} that \textbf{ICD} is true, in the sense that any individual observer in a QBist world will always assign probabilities in accordance with \textbf{ICD}, nonetheless in QBism reality is not obliged to conform to the probabilities predicted by an agent and therefore it need not be the case that \textbf{ICD} is obeyed by the actual outcomes that are obtained. 

However, this doesn't help QBism much, because although it doesn't insist that \textbf{ICD} is always true, it also can't possibly tell us that \textbf{SF} is reliably true. For QBism is very clear about the point that \emph{`measurement outcomes are personal}' and   `\emph{My measurement outcomes happen right here, to me, and I am talking about my uncertainty of them},' which entails that there can be no automatic expectation of any kind of agreement between outcomes of different observers. And since the QBist tells us that  quantum probabilities are purely subjective and do not describe or constrain anything in reality,   we cannot use those probabilities to draw conclusions about the relationships between outcomes of different observers.  Nor does the Bayesian part of QBism help with this, as by definition the QBist's coherence requirements apply only to the credences of a single observer, so they tell us nothing about the relationships between the perspectives of different observers. So although QBism does not necessarily have to obey ICD, nor is there any reason for QBists to think that \textbf{SF} is generally true, so QBist agents are also trapped in their own `island universes.'

\paragraph{Perspectives on SF}

We note that many proponents of   orthodox interpretations by and large seem to agree that their approaches are not compatible with \textbf{SF}. For example, Dascal writes that  in a neo-Copenhagen interpretation, `\emph{there need not be any causal arrow between what Alice observes and WignerA's outcome}' - i.e. we should not assume that Alice's outcome has any causal effect on the outcome of the measurement of WignerA (who plays the role of Bob in our scenario). Similarly, Fuchs tells us that according to QBism, `\emph{What we learn from Wigner and his friend is that we all have truly private worlds in addition to our public worlds.'}\cite{2010arXiv1003.5209F}. And Healey presents several no-go theorems which aim to show that, under certain assumptions, if quantum mechanics is universal than quantum measurement outcomes cannot always be objective in the sense that `\emph{(an observer's) outcome is accessible to other observers by consulting her records}'\cite{2018qtatloo}.

However, some proponents of orthodox interpretations have tried to argue that \textbf{SF} actually does hold in their approaches. For example, in the neo-Copenhagen approach of Brukner, it is argued that although in general the information about Alice's outcome (in this case, the outcome of a coin toss) is not available to Bob, nonetheless `\emph{Were he to measure in a basis that instead corresponded to Alice's coin toss (i.e. a basis whose vectors corresponded to ‘Alice having observed heads’ and ‘Alice having observed tails’), then the theorist may explain, perhaps, that his information frame grows or changes to encompass Alice's as well.}\cite{brukner2015quantum} While we agree that this seems sensible, we emphasize that it cannot possibly follow from unitary quantum mechanics. In fact in making this kind of suggestion the orthodox interpretations are illegitimately piggybacking on intuitions developed in the context of quantum mechanics \emph{with collapse}. For  if we take it that the wavefunction collapses when Alice performs a measurement, it follows that after the measurement Alice will be in an eigenstate of the form $| heads \rangle$, and hence clearly when Bob measures her he will obtain the result $| heads \rangle$. But orthodox interpretations specifically deny that the wavefunction collapses relative to Bob when Alice performs a measurement (they insist, for example, that it is still possible for Bob to observe interference between the `heads' and `tails' branch of Alice's superposition) and therefore after the measurement Alice is not in an eigenstate relative to Bob, so unitary quantum mechanics  gives us no reason to expect that  Bob's measurement will have an outcome that matches Alice's. So if the proponents of orthodox interpretations want to insist that Bob's measurement outcome will in fact match Alice's, they owe us some account of how this is supposed to come about: if all facts are relative to perspectives and measurement outcomes are just `personal experiences,' what is the justification for assuming that in the specific case where Bob and Alice use the same basis there is suddenly a guarantee of agreement between them?  Orthodox interpretations which try to insist on \textbf{SF} are trying to have it both ways: they are sneaking in some observer-independent facts which are not part of unitary quantum mechanics to ground their expectation of agreement between perspectives in certain cases, while still insisting that everything is observer-dependent and nothing needs to be added to unitary quantum mechanics.

In a similar vein, some QBists have suggested that although \textbf{SF} is not true in QBism, nonetheless we can  achieve intersubjective agreement after the fact:   `\emph{An agent-dependent reality is constrained by the fact that different agents can communicate their experience to each other ... Bob’s verbal representation of his own experience can enter Alice’s, and vice-versa. In this way a common body of reality can be constructed, limited only by the inability of language to represent the full flavor  ... of personal experience}'\cite{2020qbism}. However, we have already seen that this is not the case, because QBism can provide no guarantee that $M_A$ matches $M_B^A$, and $M_B^A$ could  represent an instance of verbal communication, e.g. Bob  asking Alice what her measurement result was. If we are really to believe that quantum mechanics is universal then it must describe verbal utterances as well as other possible methods by which Bob could `measure' Alice, and therefore verbal utterances cannot magically ensure intersubjective agreement.  Moreover, if we \emph{did} allow that verbal utterances were somehow privileged over all other types of macroscopic events, such that  that Alice and Bob can disagree about readings on measuring instruments but   they will never disagree about the content of verbal communication, then QBism would be empirically inadequate, because this would imply that experimentalists should frequently find themselves inspecting measuring instruments and then  in subsequent conversations being told that their colleagues have seen a completely different reading on the same measuring instrument at the same time - and to our knowledge this is not something which regularly happens!  Thus we must conclude that in a QBist world, Bob has no way to know whether or not the words he is hearing from Alice are the same as the words she imagines herself to be speaking; QBism puts us all within our own `truly private worlds,'   and because of its own  determined reticence about observer-independent reality it can't allow us to say anything at all about the way in which those private worlds are connected together via verbal communication or otherwise. Thus in QBism there is no way to build up a `common body of reality,' because the theory gives us no reason to expect that communication via language will work properly when observers try to communicate disagreements in their perspectives. 

Finally, some proponents of orthodox interpretations seem to accept that \textbf{SF} fails to be true in special cases, but maintain that in most circumstances it will be true. For example, after presenting his thought-experiment aiming to show that quantum measurement outcomes are not always objective, Healey observes that `\emph{the circumstances of the Gedankenexperiment in the third argument are so extreme as forever to resist experimental realization. There are no foreseeable circumstances in which the argument would require us to deny that a well-conducted quantum measurement has a definite, physical outcome. The arguments considered in this paper give us no reason to doubt the sincerity or truth of experimenters’ reports of definite, physical outcomes.}'\cite{2018qtatloo} But this conclusion gets the burden of proof the wrong way round. Healey seems to assume here that   observers are guaranteed to agree on measurement outcomes except in the special cases where it is possible to prove a no-go theorem to the effect that intersubjective agreement is impossible: but the orthodox interpretations specifically deny the existence of any external reality or structure which could coordinate the perspectives of different observers, so they are not entitled to take intersubjective agreement for granted in any scenario, even if there aren't any no-go theorems for that scenario.  No-go theorems of this kind, if accepted, simply demonstrate in a more explicit way that the orthodox interpretations are committed to failures of intersubjectivity, but there is no reason to think that the failures of intersubjectivity will be confined to the cases described by the no-go theorems.

 \paragraph{Decoherence}

Some proponents of orthodox interpretations believe that decoherence has no foundational significance; for example, in QBism, `\emph{decoherence does not come conceptually before a “selection,” but rather is predicated on a time $t = 0$ belief regarding the possibilities for the next quantum state at time $t = \tau$ ... In this sense the decoherence program of Zeh and Zurek ... regarded as an attempt to contribute to our understanding of quantum measurement, has the story exactly backward.}' On the other hand, some proponents of orthodox   interpretations seem to suggest that the  problem of intersubjectivity can be solved by decoherence. For example,   Brukner uses an approach based on a `Q-function' characterising probabilities for measurements of coarse-grained variables\footnote{The Q-function is an approach to classicality which has some similar features to decoherence, although it is not dynamical in character and thus differs conceptually in various ways.} to argue that for measurements on macroscopic systems `\emph{the Q-function before and after a coarse-grained measurement is approximately the same. It therefore becomes possible for different observers to repeatedly observe the same macroscopic state. The result is a certain level of intersubjectivity among them.}' 

However,   Brukner's approach, like the others we have considered, regards quantum states as relative to an observer, which entails that when Alice  performs a coarse-grained measurement, the macroscopic state of the system becomes definite relative to Alice, but relative to other observers this interaction simply leads Alice to become entangled with the relevant system as in the example we saw earlier. So when Bob performs another coarse-grained measurement on the same system, he will find a definite macroscopic state, but since this approach adds nothing to unitary quantum mechanics it cannot provide us with a  guarantee that the macroscopic state found by Bob is the same one found by Alice. So although both observers experience a stable macroscopic reality, there is no reason to expect that both observers experience the \emph{same} macroscopic reality. Similar issues would be expected to arise within other orthodox interpretations, and therefore it seems that decoherence does nothing to solve the problem of intersubjective agreement in orthodox interpretations.

 \section{Accessing Other Perspectives \label{OM}}
 
 Now, the  `ontology of island universes,' might seem unappealing - for example Pienaar objects that `\emph{This proliferation of disjoint universes is not motivated by observations, nor does it serve any explanatory purpose; it therefore seems to be (paraphrasing Wheeler) ontological ‘excess baggage’. Moreover, it is difficult to see what ‘objectivity’ could possibly mean in such a universe.}'\cite{2021quintet} But by and large the proponents of  orthodox interpretations seem willing to accept the failure of intersubjectivity. Indeed, they regard the notion that science describes an external reality on which we could hope to reach intersubjective agreement as a naive idea that must be discarded; ref \cite{QBismintro} criticizes it as  `\emph{a profound misconception in our general view of science, which led us into major confusion in the 20th century,}' and ref \cite{cuffaro2021open} suggests that `\emph{Orthodox interpreters can counter ... that the ideal of an observer-independent reality is not methodologically necessary for science and  ... modern physics (especially, but not only, quantum theory) has taught us ... that there is a limit to the usefulness of pursuing this ideal.'}
 
 But these responses, we suggest, misunderstand   the issues at stake. The problem with the ontology of island universes   is not just that it is unappealing or in tension with our na\"{i}ve classical ideas; the problem is that intersubjectivity plays a crucial role in the process by which we arrive at and confirm scientific theories, so if intersubjectivity fails, the status of quantum mechanics as empirically confirmed is endangered. And if quantum mechanics can no longer be regarded as empirically confirmed, then evidently we will have no reason to attach any credence to an orthodox interpretation, or indeed to any interpretation of quantum mechanics at all. Thus the whole approach starts to look self-undermining: no adequate interpretation of quantum mechanics can have the consequence that we should not believe in quantum mechanics in the first place, so this is an issue which should be treated with the utmost seriousness by the proponents of  orthodox interpretations. 
 
Intersubjectivity is important to the scientific process because science is a collaborative endeavour: typically we put our faith in a theory based not only on our own observations but also the observations that have been made and reported by a multitude of other scientists across the globe. For example, the proponents of the orthodox interpretations clearly have a high level of confidence that quantum mechanics is right, and yet many of them are theoreticians who have not spent much time performing laboratory tests of quantum-mechanical effects, and therefore their own observations certainly do not provide enough data to empirically confirm quantum mechanics as a whole. Even a physicist more experimentally inclined who has engaged in a variety of experimental tests of quantum mechanics will necessarily rely on the testimony of others to verify that  quantum mechanics has worked in roughly the same way across most of the Earth for at least the last 100 years, and to reproduce her experiments to make sure that she is using correct experimental technique and her observations are not just some kind of fluke. 

Obviously, even in classical physics  intersubjectivity does not hold with complete reliability. Other scientists could  lie about their results; observers could have visual impairments or hallucinations or false memories which cause them to disagree about the result of a measurement. But nonetheless in this case there is no barrier \emph{in principle} to intersubjectivity: the observations of different observers are connected to each another by the unique observer-independent reality which they are all trying to observe, so   provided that    everyone is honest and everyone's perceptual systems and memories are working correctly, everyone will agree on the results of measurements and other such macroscopic events.  Moreover, classical physics tells us that it is possible to verify the accuracy of our observations by comparing notes with other observers, so if something has gone wrong it is  straightforward to find out and repeat the experiment.

The failure of intersubjectivity in orthodox interpretations of quantum mechanics is significantly more severe, as even in the ideal case where everything is working perfectly, observers may still disagree.   And we have seen that not only will observers disagree in such cases, in general they \emph{will not even realise that they disagree.}  So for example if Alice and Bob in the scenario described above are both trying to use the outcome of Alice's measurement on $S$ to confirm a scientific theory, both of them will add different pieces of data to their set of recorded observations and will proceed with their scientific labours in blissful ignorance of the fact that they're both working with different data. And presumably, if they repeat this kind of experiment multiple times, it seems likely that they will ultimately come up with different scientific theories to match their different datasets - but any time they try to have conversations about the theories they have come up with, \textbf{ICD} tells us  that both will always be convinced that the other is agreeing with them!\footnote{As we observed earlier, \textbf{ICD} does not necessarily hold in QBism so QBism doesn't guarantee that observers will always perceive other observers as agreeing with them - but as we noted, QBism also cannot provide any reason to think that verbal communication is accurately conveying the content of experience, so the same problem arises.}

Furthermore, intersubjectivity is actually \emph{more} important in orthodox interpretations than it is in the case where we regard quantum mechanics as a description of an observer-independent reality. For orthodox interpretations tell us that quantum mechanics is  specifically \emph{about} the structure of a network of perspectives, not about external reality, so when we are trying to obtain empirical confirmation for the theory, the only kind of information that matters is information about the structure of this network of perspectives. Moreover, clearly information about just one perspective will not tell us much about the features of the whole network, and yet we will shortly see that it is important for orthodox interpretations to maintain that quantum mechanics is a theory about the whole network, not just a single perspective. So in order for quantum mechanics to have any empirical confirmation in this context, it is vital that we should have access not only to our own perspective but also to the perspectives of some other observers - exactly what the orthodox interpretations apparently forbid us from having!

 Orthodox interpretations  thus face a stark dilemma:    either there are no grounds for expecting intersubjective agreement between observers about measurement results, in which case it will be impossible to gather the kind of evidence that is needed for quantum mechanics to be empirically confirmed, or there \emph{are} grounds for expecting intersubjective agreement between observers in at least some relevant subset of cases, in which case it's simply not true that quantum mechanics does not describe an observer-independent reality  - the observer-independent reality is precisely what one gets from combining the perspectives of observers in the cases in which they typically agree.    It is therefore clear that  intersubjectivity cannot be discarded too lightly: if we \emph{are} going to postulate that intersubjectivity fails in the way posited by these orthodox interpretations, it is vital that we should be able to offer  a good story about how it is that quantum mechanics could come to be empirically confirmed in a world like this.

\subsection{Confirmation} 
 
 Let us now take a closer look at how the process of empirical confirmation might work in the context of an orthodox interpretation. We begin with the idea that in order for me to be able to arrive at and  empirically confirm scientific theories, at minimum I must be able to   to look at a measurement outcome and use that observation to update my beliefs about the world as a whole - that is, it must be the case that when I make an observation, I will sometimes update my beliefs in such a way that the probabilities I assign to various different possible worlds will change. Moreover, we take it that a defining feature of a scientific theory is that it is generalisable - it does not just describe a set of observations, but also allows us to extrapolate beyond the observations that we have used to arrive at it. So in order for me to  be able to empirically confirm scientific theories, it must be the case that when I make an observation, I will sometimes update my beliefs about something other than the observation I have just made - i.e. the probabilities I assign over events in parts of the world outside of the region of my observation must change. 
 
 Now, in the case of orthodox interpretations this is already complicated by the fact these approaches typically insist that `\emph{there is no description of the universe “in toto,” only a quantum-interrelated net of partial descriptions.}'\cite{1996cr} This makes it a little hard to understand what exactly we are trying to confirm when we seek empirical confirmation within an orthodox interpretation. However, these approaches are mostly intended as realist views and thus they do postulate something real: in general they agree that we have a set of observers whose experiences are subject to reliable probabilistic relationships as predicted by quantum mechanics.  So we will take it that in this context the role of a `possible world' is played by a possible \emph{network of perspectives}, i.e. a set of perspectives together with the observations occurring within each of the perspectives. We emphasize that the orthodox interpretations tell us that quantum mechanics is universal, i.e. it correctly describes all the perspectives in the network, so what we are trying to confirm is not just the content of a single perspective but the regularities obtaining in many different perspectives.  Thus in this section we will suppose for the sake of argument that we believe ourselves to live in a world in which there are no  observer-independent facts (except possibly ineffable ones), but there are some facts about the regularities that obtain within different perspectives, and we are trying to confirm the hypothesis that the regularities that obtain across the whole network of perspectives are approximately the ones specified by quantum mechanics.  We will then ask what empirical confirmation could look like in such a scenario.
 
 We pause to note that the role of confirmation may seem less clear in the case of a view like QBism, which holds that quantum mechanics is normative rather than descriptive - i.e. it simply tells agents how to assign probabilities. One might wonder if it makes sense to seek empirical confirmation for a purely normative theory, and indeed, Fuchs has argued that as far as possible we should seek to derive the mathematical framework of quantum mechanics from pure information-theoretic principles. But plausible as those information-theoretic principles might be, they are not tautologies, so we still have to rely on empirical evidence  to find out if they hold in our actual world - for example,  although the Born rule may have some Bayesian justification, empirical input is also needed to arrive at it. Indeed, if we were to say that quantum mechanics actually has nothing whatsoever to do with the empirical evidence that has led us to arrive at it, it would be an amazing and inexplicable coincidence that the empirical evidence has led us to a  theory which the QBists believe to be correct, even though that theory does not actually have any connection to the evidence! So it  seems that the QBists must accept that quantum mechanics is at least to some degree susceptible to empirical confirmation. And indeed, most QBists seem to accept that their normative principles are `\emph{empirically based (as quantum theory itself is)}'\cite{2010arXiv1003.5209F} so  we think it is reasonable to demand that QBism   should allow for empirical confirmation, in the sense that we should be able to use empirical observations to gain support for the QBist's proposed empirically motivated norm of rationality.
 
 \subsection{Minimal conditionalisation}

One common model for belief-updating based on empirical evidence  is minimal conditionalisation, which ‘\emph{simply amounts to setting one's credence to zero on the proposition whose falsity one has just learnt, and renormalising.}’\cite{Greaves2006-GREOTE} For example, if we perform an experiment at time t and observe outcome E, we learn that the proposition ‘E does not occur at time t,’ is false, and we should therefore eliminate all possible worlds in which E does not occur at t. But it is hard to see how we could apply minimal conditionalisation in the context of  an orthodox interpretation, because we simply don't seem to be dealing with the right sort of objects. When I perform an experiment at time t and observe outcome E, I learn that the proposition ‘E does not occur at time t,’ is false \emph{relative to me}, and therefore I can eliminate all possible descriptions of the world \emph{relative to me} in which E does not occur at t, then renormalise my credences across descriptions of the world \emph{relative to me}. There does not seem to be any point at which I can drop the `relative to me' modifier and say that I have learned something about the network of perspectives as a whole: for \textbf{ICD} entails that even when other observers report their results to me, I will hear the content of their report \emph{relative to me}, which may be different from what they themselves imagine the report to be saying!  So it is clear that minimal conditionalisation is not adequate for empirical confirmation in the kind of world described by an orthodox interpretation, since it will never allow agents to update their beliefs about regularities obtaining in perspectives other than their own.

\subsection{Centered Worlds} 

Now, one might hope that these problems could be addressed by moving to an alternative account of belief-updating. For example, Lewis offers an approach designed to deal  with the case of self-locating beliefs, i.e. beliefs about where or when one is located within a given possible world.  He postulates a set of ‘centered worlds,' where `\emph{roughly, a centered world is an ordered set of a possible world and a perspective within it,}'\cite{10.2307/41426893}  and he claims that Bayesian conditionalisation can be applied straightforwardly to self-locating beliefs by replacing possible worlds with centred worlds\cite{Lewis1979-LEWADD}. On this approach, the credence assigned to a possible world is the sum of the credences assigned to all of the centred worlds associated with it, so the credences assigned to a possible world as a whole may be altered by a change in our credences for the centred worlds associated with that possible world, and therefore learning purely self-locating information can alter our beliefs about reality as a whole. 

We might hope to make use of this strategy in the context of orthodox interpretations by defining a centered world as a perspective relative to a certain observer. That is, let a centered world be an ordered set of a network of perspectives (which plays the role of a `possible world' in the orthodox interpretation context) and a `perspective' (in the sense of an orthodox interpretation, i.e. a set of facts relative to an observer or similar). Then when I make an observation I can rule out all the centered worlds which are incompatible with my observations,  which may result in changes in the number of centered worlds associated with some possible networks of perspectives, and thus overall the probabilities I assign to various different networks of perspectives may change, making it possible to perform empirical confirmation in the orthodox context after all. 
 
However, note that we can used centered worlds to arrive at meaningful empirical confirmation for scientific theories only if we start out with some priors such that in some cases, our priors assign \emph{different} probability distributions over the content of the `rest of the world' conditional on subjectively different centred worlds. That is to say, let the `rest of the world' be the content of some possible world minus the content of  some particular centered world within that possible world. Let us  form equivalence classes of `subjectively identical' centered worlds, i.e. centered worlds or perspectives which belong to different possible worlds but are indistinguishable from the point of view of the observer to whom they belong;  when we make an observation, we will rule out all the equivalence classes of centered worlds which are inconsistent with our observation and renormalise. Thus if our priors assign the same probability distribution over the `rest of the world' conditional on every possible equivalence class of subjectively identical centered worlds, if follows that the probability distribution we assign over the `rest of the world' will not change during this updating process. Thus under these circumstances, an observation does not allow me to update my beliefs about anything other than the observation that I have just made: I learn that the actual world contains a centered world with certain subjective features, but I learn nothing at all about what other centered worlds feature in the actual world or how these centered worlds are related, so I can't obtain confirmation for any beliefs about the world which extrapolate beyond my direct observations, and as we have already noted, in order for me to be able to arrive at and empirically confirm a theory like quantum mechanics it must sometimes be possible to extrapolate beyond direct observations. 

In ordinary cases of empirical confirmation,  our priors do indeed assign different probability distributions over the content of the `rest of the world' conditional on subjectively different centred worlds. For example, this is often achieved by choosing priors which assign a higher weight to possible worlds which look `simple' in some way, e.g. worlds in  which regularities typically persist across time. But we do not seem to have this option in  a setting where we are not allowed to postulate any knowable observer-independent facts.  In particular, we saw in section \ref{wide} that orthodox interpretations do not allow us to postulate any that there should be any systematic relations between  observations made by different observers, so it would seem that if we believe ourselves to be in a universe of the kind described by an orthodox interpretation, we cannot  select prior probabilities which favour networks of perspective exhibiting certain sorts of relations between perspectives:  we can't select priors which encode beliefs like   `regularities typically persist across different perspectives'   since those beliefs, if true, would constitute knowable, observer-independent facts. Indeed it's hard to see how we could formulate any prior probabilities at all under this limitation (even the uniform distribution is not really a neutral choice, as it still involves choosing \emph{some} measure over the relationships between perspectives).

\subsection{Adding more assumptions}

It seems, therefore, that neither minimal conditionalisation nor Lewisian centered-world conditionalisation is going to lead to any meaningful empirical confirmation in the context of an orthodox interpretation. Of course, it is possible that  a new account of confirmation could be proposed specifically for this scenario, and indeed we would invite the proponents of orthodox interpretations to take up this challenge. But there is an overarching problem which seems likely to inflict any account of confirmation in the context of orthodox interpretations:  if we are to understand how quantum mechanics (or any other scientific theory) could be empirically confirmed in this context, we must be provided with some bridging principle telling us how to get from the facts about events within a single perspective to facts about the rest of the network of perspectives, but because these interpretations do not allow us to postulate knowable, observer-independent relations between perspectives no such bridging principle can exist. Thus it seems that in these kinds of approaches there is simply no way to get outside an individual perspective and say something about reality which is not relativized to one's current perspective.

Proponents of orthodox interpretations might hope to solve these problems by slightly relaxing the requirement that there are no knowable observer-independent facts, in order to allow the assumption (A1) that most or all of the perspectives in the actual network of perspectives exhibit similar kinds of regularities, even if they disagree on the specific outcomes of certain measurements. This assumption could perhaps be grounded on the principle that I am not in any way special as a physical system, and therefore the probabilistic relationships exhibited in my own evidence should also hold for other observers. The proponent of orthodox interpretations would then presumably hope to employ this assumption in the same way as in ordinary cases of empirical confirmation we typically employ an assumption to the effect that  (A2) regularities persist through time. 

However, these two assumptions are not analogous. For  assumption (A2) is susceptible to empirical enquiry: we do experiments to test \emph{which} regularities persist, we test out systems   in new regimes to see if the usual regularities still hold, we try to come up with explanations for regularities in terms of common causal factors or underlying mechanisms. Moreover, observing regularities persisting in the short term provides  some grounds for the assumption that they will continue to persist - at least we know that these regularities  \emph{sometimes} persist. So although we do ultimately have to make an unprovable assumption that the future will be in some ways like the past, the parameters of that assumption are not ad hoc: we use empirical evidence to decide  the specific ways in which we expect the future will be like the past.

But in the case of assumption (A1) no such empirical enquiry is available. We will never have access to any perspective other than our own, so we can't do experiments to test out the ways in which other perspectives are similar to our own, or check to see if there are regimes in which those similarities break down, or offer any explanation in terms of common causal factors or an underlying mechanism. And since we have never accessed \emph{even one} other perspective, we don't have any empirical grounds at all for our assumption of similarity between perspectives - we don't even know that different perspectives \emph{sometimes} look similar. So necessarily, the parameters of the assumption that other perspectives are like our own are completely ad hoc: no empirical evidence can possibly tell us anything about the ways in which different perspectives are similar to ours. So how do we know where to stop with our assumptions about similarity of perspectives? The principle that `I am not special' does not  tell us much: after all one way to satisfy this principle would be to  say that all perspectives are identical, and another would be to say that they are all completely random. In fact, obviously where proponents of the orthodox interpretations would like to stop is precisely at the limits of quantum mechanics: we are supposed to take it that all perspectives obey quantum mechanics, but they differ in other ways. Yet there seems to be no real justification for this assumption, so the approach taken by the orthodox interpretations   amounts to transferring the conclusions reached in our original enquiries about the persistence of regularities over time to a completely different question about similarities of perspectives, even though there is no reason to think the former has any relevance to the latter.

Of course,  it will be responded that even in classical physics we have no way to  be sure that other people's subjective perspectives are like our own. But the  crucial difference  is that  orthodox interpretations, unlike classical physics,  are \emph{about} perspectives: they tell us that quantum mechanics is nothing more than a characterisation of a network of perspectives associated with different observers. In the case of classical physics, because our goal is to empirically confirm  a theory which is about the structure of mind-independent reality, it doesn't really matter that we don't know for sure what subjective experiences other people are having:  in order to perform empirical confirmation  we only need to assume that our \emph{own} subjective experiences gives us information about the structure of mind-independent reality. But in order to confirm a theory which is explicitly about a network of perspectives,   our experiences must give us information about the network of perspectives, including information about perspectives other than our own. Yet the orthodox interpretations specifically tell us that our  experience cannot ever give us such information, and thus they do not allow that we can have empirical access to the very objects (perspectives) which are the subject of the theory - or rather, we can have access to only one such object, which clearly will not provide us with enough data to confirm a universal theory about the  properties of these objects.

To make matters worse, not only do the orthodox interpretations hinder us from making hypotheses about how different perspectives relate to one another, they often don't seem to have the means to tell us which perspectives will exist in the first place. The ontology of an orthodox interpretation consists of a set of observers together with their perspectives: in RQM any system whatsoever can be an `observer,' whereas in QBism and most Copenhagen and neo-Copenhagen views it seems that only conscious beings count as observers.  But \emph{which} observers are there? In the usual case where we have an observer-independent objective reality, we can look at some observer-independent theoretical model of reality and pick out the objects in it which play the role of observers. But in an orthodox interpretation there is no observer-independent reality, or at any rate observer-independent reality is  ineffable, so where should we get our set of observers from? My own perspective will presumably include a a set of objects which might count as observers, but it is an important part of the motivation for these views that no observer is special, so the set of observers can't be defined by which observers exist relative to me - if observers can only have their existence in virtue of existing in a set of facts relative to some other observer, it would seem that we would end up with an infinite regress which will not give rise to a well-defined set of observers. So does every logically possible observer exist, seeing every logically possible description of reality, and thus every possible set of relative frequencies? Or is some set of observers simply picked out of thin air? Either way, if we don't have a well-defined set of observers it seems difficult to even formulate the claim that most perspectives exhibit similar kinds of regularities  - in order to make sense of this claim we'd have to say something about the proportion of observers who see relative frequencies which exhibit the same kind of regularities, and we can't make sense of such a claim without first having some concrete idea of what our set of observers looks like.

\section{Persistence over time \label{past}} 

One possible response   to our concerns would be to argue that although  an observer in a world described by an orthodox interpretation may not be able to learn about observations made by others, she can at least perform empirical confirmation using her own observations and her memories of the measurement results that other observers have reported to her (while accepting she can't be sure that her memories of those reports correspond to what the other observers experienced themselves as perceiving) and thus she can at least confirm that her own experiences are as predicted by quantum mechanics.   Evidently the theory that this observer can confirm is of much more limited scope than quantum mechanics as it is usually understood, but nonetheless it would seem that there is \emph{something} that she can confirm, and perhaps one might argue that this is enough to justify the practice of empirical science.

However, this argument relies crucially on the question of whether or not  `observers' can be understood to persist through time in orthodox interpretations. If observers do persist and the set of facts  relative to a given observer remains the same throughout their existence, then indeed these observers may be able to empirically confirm quantum mechanics as a description of their own experiences, but this will clearly not work if we are forced to say that an  `observer' exists only at a single instant, because then later versions of the same person must be treated as distinct `observers'  who may have a completely different set of facts.

 And unfortunately for the proponents of orthodox interpretations, it seems quite difficult for them to insist  that observers persist over time. For in order to do this, the orthodox interpretation would have to provide some criterion of identity over time so we can track which temporal parts of a continuant are supposed to count as the same observer and thus have the same set of relative facts, and it seems hard to do this in any precise way. Do I cease to be the same observer when all the cells or particles in my body have been replaced? Perhaps only the particles in my brain are relevant, or the particles in some particular part of my brain? What about `ship of Theseus' examples\cite{Jansen2011-JANTSO-4}, where an observer is slowly deconstructed and reassembled elsewhere: when does the observer in the process of deconstruction cease to be identical with the original observer, is the reassembled observer ever identical with the original system, and  can there be some time at which \emph{both} observers are identical to the original? 
If we want to claim that there is some objective notion of continuity that provides answers to these questions, we would essentially need to postulate something akin to a disembodied soul which tracks `the same person' over time, and this will surely be unacceptable to most proponents of orthodox interpretations, particularly those motivated by broadly physicalist and naturalist intuitions. But if we accept that these questions are essentially just semantic ones to be answered by appeal to our linguistic conventions, then surely the account of the content of   reality given by an interpretation of quantum mechanics cannot depend in any crucial way on the answers to them. 

QBism might perhaps be more friendly to the `disembodied soul' approach, as by construction QBism represents  `\emph{a radical departure from the strictly physicalist ontology endorsed by nearly all other quantum interpretations,}.' However, note that according to QBism, what it is to be an   `observer' is to be a decision-theoretic agent whose beliefs are regulated by a principle of coherence. And the principle  of coherence   does not allow us to make   comparisons between observers at different times: `\emph{Coherence is a condition about an agent’s} current beliefs, \emph{including her beliefs about her future probability assignments.}' Thus QBism does not postulate any direct connections between the beliefs of observers at different times: it demands that an agent's current beliefs about what beliefs they will hold in the future must be coherent, but it does not allow any diachronic rationality constraints.  This means that QBists  cannot postulate any prima facie connections between observers at different times,  meaning that  QBism likewise cannot offer any guarantee that an agent's beliefs about what has happened in the past will remain stable over time.

We emphasize that we would not, under normal circumstances, expect a scientific theory to provide an answer to questions about personal identity over time; that kind of philosophical question is usually outside the realm of science. But by making `observers' so central to their approach, orthodox interpretations impose on themselves a mandate to tell us in precise terms what they mean by an observer, because that is a crucial step in defining the empirical content of the theory. The criteria for tracking an `observer' in the technical sense used by the orthodox interpretations need not coincide with the criteria which we would naturally employ to track personal identity over time in the philosophical sense, but \emph{some} criteria must certainly be provided if proponents of these interpretations want to insist that we can at least use our own memories for empirical confirmation; and our concern is that it seems far from clear that there is  \emph{any} acceptable set of criteria available for this purpose, even if we don't demand that these criteria also track the common-sense notion of personal identity.

Now, one might hope to avoid the problem by saying that the  the `observers' with which an orthodox interpretation is concerned are all individual fundamental particles -  that is to say,  all `perspectives' must actually be relativized to a single particle - and therefore tracking identity over time just comes down to following the trajectories of individual fundamental particles over time. This is not an option within orthodox interpretations like QBism which specifically require that their observers are conscious beings, but it might be viable within something like RQM, which tells us that every physical system can be an `observer' in a general sense.  However, we probably do not want to make an interpretation of quantum mechanics contingent on the claim that there exist fundamental particles following continuous trajectories through spacetime, since this would likely cause problems when we came to apply the interpretation to field theory, which tells us that particles are simply excitations in fields and thus there is  often no meaningful notion of particle identity across time\cite{pessa2009concept}. Moreover,  the   solution of the measurement problem offered by orthodox interpretations depends crucially on the claim that   there is some unique set of facts relative to each (human) observer - that is, a measurement always has a unique outcome relative to the person who performs the measurement. This means that a person cannot simply be regarded as the composite of all of the particles in their brain,  since we have already seen that there can be no guarantee that the descriptions relative to all of these different systems will agree, and therefore their composite would not give rise to a well-defined perspective. Thus we do need to allow that conscious minds  count as `observers,' which are distinct from all of their individual particles; and since there is no   unambiguous way to identify such things across time, it would seem that if the orthodox interpretations wish to maintain their clean, unambiguous picture of facts as relative to individual observers, they had better maintain that such  observers are defined only at a single time.  And in fact, many proponents of orthodox interpretations do indeed seem to accept that observers do not persist over time. For example, Ruyant writes that in RQM, after a measurement the observer `\emph{has just became a new observer with a new perspective, relative to which objects have different properties.}'\cite{Ruyant2018CanWM} 

Indeed, quantum mechanics itself provides concrete examples of cases where the perspective of an observer cannot remain stable over time. For example, ref \cite{2021nogt} presents  a no-go theorem shows that in a Wigner's friend experiment we cannot assign a joint probability distribution across the observations of the friend before and after Wigner's experiment unless we give up the assumption that unitary quantum mechanics is universal or the assumption that the joint probability of the friend’s perceived outcomes  has a convex linear dependence on the initial state of the system qubit. It seems likely that most proponents of orthodox interpretations would wish to uphold these two assumptions, since the universality of quantum mechanics is one of the basic principles of an orthodox interpretation, and consequently they must accept that at least in this case it isn't possible that the observer retains a consistent perspective over time. Of course examples like the Wigner's friend experiment are presumably quite rare, but nonetheless they serve to underline the point that the orthodox interpretations have no sensible way to define the continuity of a perspective over time.

\subsection{Relative Frequencies} 

If orthodox interpretations cannot justifiably postulate that `observers' in their technical sense persist over time and thus retain the same set of relative facts, this will clearly have serious consequences for empirical confirmation, particularly with reference to  the probabilities which  feature in quantum mechanics and which form an important part of its empirical content.\footnote{There are a variety of positions on the nature of probability in orthodox interpretations. Because they all insist that quantum mechanics is complete, they all agree that the probabilities are `irreducible' in the sense used by Brukner\cite{brukner2015quantum}, i.e. there are no hidden variables in the theory such that conditioning on these variables would allow us to predict quantum measurement outcomes more precisely than the Born rule. However, there are some differences of opinion about the status of these irreducible probabilities: in QBism they are purely subjective probabilities: `\emph{The starting idea of QBism—which originally stood for “quantum Bayesianism”—is that the probabilities delivered by quantum theory are to be understood as degrees of belief along the lines of the subjective Bayesian approach to probability}'\cite{pittphilsci16382}, but in other contexts they may be regarded as objective, as for example in  Dorato's proposal for a dispositional interpretation of the probabilities in RQM\cite{Doratodisp}.} In particular, we note that proponents of the orthodox interpretations often wish to characterise probabilities as referring specifically to the results of measurements yet to be made. But quantum mechanics  obtains its empirical confirmation in large part from observations of relative frequencies - e.g.  the pattern of dots on the screen from a two-slit experiment, which constitutes a visual representation of relative frequencies.  A single incidence of a particle on the  screen tells us virtually nothing; it is the picture built up from recording many incidences on the screen which allows us to draw conclusions about the interference processes taking place. So if the theory is to have empirical confirmation, clearly the probabilities assigned by the theory must be related not only to the results of future measurements but also to relative frequencies observed in the past: it must be the case that the structures postulated by the theory have some part in the explanation for why we have seen these relative frequencies and not others. 

Now, in an interpretation of quantum mechanics which postulates a stable, intersubjective macroscopic reality,   if measurement outcomes typically \emph{occur} with frequencies matching the Born rule probabilities it is reasonable to suppose that subsequent memories and records of those measurements will exhibit relative frequencies which approximately match the Born rule probabilities. But we have seen that in orthodox interpretations it can't be taken for granted that an observer's perspective will be stable across time. So in order for quantum mechanics to receive confirmation from relative frequencies within an orthodox interpretation, we must say something more about the relationship between memories and records and the probabilities postulated by the theory.   
Thus let us once again suppose for the sake of argument that   we believe ourselves to live in a world in which there are no observer-independent facts (except possibly ineffable ones), but there are some observer-dependent facts about the measurement outcomes that appear in different perspectives, and we will imagine that we are trying to confirm the hypothesis that the correct probability distributions to describe at least the events featuring in our own experiences are the ones specified by quantum mechanics.  Since we can't assume that our memories represent what actually happened in the past, in order to make progress on confirming this hypothesis we have to make a decision about  the status of memories and records in this kind of world; in the following two sections we consider two possible options. 

 We pause to note that relative frequencies may seem to pose a particular problem for QBism, since it denies that there are any objective facts about the probabilities for quantum measurement outcomes, even relative to a single observer: `\emph{the individual outcome of a quantum measurement is random and lawless}'\cite{2010arXiv1003.5209F}. But as we have already seen, a hardline QBist position maintaining that quantum mechanics has nothing whatsoever to do with the empirical evidence  would seem very hard to accept, so presumably the QBist must hold that relative frequencies play some part in the justification for the normative principles that they associate with the quantum formalism. It is not entirely clear to us how relative frequencies can provide support for a theory which explicitly tells us  that all probabilities are purely subjective, and Timpson also raises this as a significant concern: `\emph{if there are only subjective probabilities ... why does gathering data and updating our subjective probabilities help us do better in coping with the world? ... Why, that is, should one even bother to look at data at all?}' But we will leave that for the QBists to explain; for now, we will simply take it that  in QBism, and in all other orthodox interpretations, the reasons for believing quantum mechanics are at least partly based on relative frequencies, so it is important that we can attach meaning to the relative frequencies that we have observed.

\subsection{Memories as measurements}

One possibility is that  consulting memories and/or records should be understood as a  \emph{measurement} on some kind of physical register (e.g. the human brain), and hence the theory we are trying to confirm can be understood as prescribing a probability distribution over measurements on memories and records.  Orthodox interpretations do not allow us to make the assumption that these memories and records  are an accurate reflection of what `really happened,'  since what `really happened' will in general be relative to a different observer  - i.e. they will either be relative to a past version of ourself, or they will be relative to some other observer who performed the measurements and passed on the records. Moreover,   we can only arrive at relative frequencies by consulting measurement and records, as we cannot directly observe relative frequencies in a single measurement. Even if we construct some kind of joint global measurement which is supposed to measure the relative frequency of some property in some ensemble, a single measurement is not enough to tell us that the result of this joint global measurement has anything to do with the relative frequencies that we would have observed if we had measured the members of the ensemble individually - that expectation could only be based on the fact that quantum mechanics tells us how joint global measurements are related to measurements on individual systems, but without independent access to the individual measurements we have no reason to have any faith in that part of quantum mechanics. 

Thus it seems that under such circumstances, any theory we come up with on the basis of observed relative frequencies can only be regarded as a theory of \emph{the results of measurements on memories and records}! For we only ever have access to  the sequence of outcomes recorded in our memories and records, not the actual sequence of  outcomes which originally occurred in the perspectives of the observers who performed the measurements, and in the context of an orthodox interpretation  we can't take for granted that these sets are the same. Thus if we are to use our memories to empirically confirm quantum mechanics, we must assume  that (A3) the set of measurement outcomes derived from a measurement on our memories and records  will typically exhibit  similar \emph{relative frequencies} to the set of outcomes that originally occurred in the perspectives of the observers who performed the measurements, even though for any specific measurement outcome the recorded value may be different to the value that originally occurred.  But what grounds could we possibly have for this assumption? There is no way we can ever check assumption (A3); it is simply drawn verbatim from standard quantum mechanics, but remember that on this construal quantum mechanics has empirical confirmation only as a theory of memories and records, and we therefore can't use it to infer (A3), since we have no evidence that quantum mechanics is a correct theory of anything \emph{other} than memories and records. Thus if consulting memories is to be understood as a measurement which is governed by the theory we are trying to confirm, it seems we cannot arrive at a theory which says anything about what we will see the next time we do a measurement: we may be entitled to draw the conclusion that, if we perform a sequence of measurements and then subsequently inspect the records, we will find that these new relative frequencies approximately match the ones we observed earlier, but we have no grounds to make any predictions about what we will see at the time of the measurement.

 Moreover, on this construal we don't even have enough data to justify coming up with a theory which at least applies to measurements on memories and records. For we will only ever have access to the result of \emph{one} measurement, i.e. the measurement we have just made on our memories/records of the past. We can't measure our memories now and then later measure our memories again and compare the two results, because we don't have access to the result of the earlier measurement except via the part of the outcome of the later measurement which pertains to our memory of the earlier measurement, and as usual there is no guarantee that the two agree. So we can never have access to a sequence of outcomes of distinct, independent measurements on our memories and records, and therefore we can't obtain any relative frequencies in order to arrive at a theory in the first place.
 
Of course, one might object that this predicament is not unique to the orthodox interpretations - any scientific theory  must  in some sense be a theory of  memories and records rather than the events in and of themselves, since theories can only ever be arrived at on the basis of memories and records. But normally when we do science we  start from the assumption that there exists a stable intersubjective macroscopic reality and therefore we are free to assume that (A4)   our memories and records are mostly accurate reflections of what actually took place in this stable intersubjective macroscopic reality, and it is this assumption which allows us to regard ourselves as having access to the result of more than one distinct measurement. Of course, the proponents of  orthodox interpretations might try to insist that assumption (A4) is  no different to assumption (A3) and thus the orthodox interpretations are no worse off than any other scientific theory. For example, this seems to be the approach taken by some proponents of QBism: `\emph{all these data and observations are obtained through experimental conditions that are without exception instrumentally-mediated which in turn means that we are already experiencing an interpreted reality.}\cite{2020qbism} But in fact (A3) and (A4) are not equivalent, because   assumption (A4) is independent of any particular theory we might be trying to confirm: it is based on a pre-existing philosophical commitment to the existence of an observer-independent reality, which allows us to infer that when everything is working properly our memories will reflect that observer-independent reality and thus will accurately reflect what actually happened in the past. That is to say, if we assume the existence of an observer-independent reality then the relative frequencies exhibited in memories can be expected to match the relative frequencies exhibited in the actual events because the latter are understood to exist in an observer-independent sense and therefore they can plausibly be the direct cause of the former. Whereas in the case of the orthodox interpretations  we are no longer allowed to postulate an observer-independent reality (or at best we can only postulate an ineffable and unknowable one) and therefore there can be no pre-theoretic reason to think that relative frequencies exhibited in memories match the relative frequencies exhibited in the actual events, since the latter do not exist in any observer-independent sense and therefore they cannot be the cause of the former. The only justification for assumption (A3) comes from appealing to QM itself,  but we cannot arrive at QM in the first place without making the assumption (A3), so this looks like a vicious circularity. Evidently (A3) has just been tacked on ad hoc   because  proponents of orthodox interpretations wish to be able to arrive at standard quantum mechanics, but really the scientifically responsible thing to do here would be to come up with a theory which describes only the  results of measurements on memories and records and which is silent on the relationship between memories and actual events - and that theory would certainly not be standard quantum mechanics.

Now, the proponent of an orthodox interpretation might try to justify their use of (A3) as some kind of bootstrapping procedure: one starts by assuming (A3), one arrives at the theory of QM on this basis, and then one uses the theory to justify the assumption (A3). But care must be taken with this kind of reasoning, because obviously I could `justify' more or less any assumption I liked by first making the assumption, then  using the assumption  together with some evidence to infer some conclusion, and then observing that the conclusion implies the assumption that I used to derive it. So bootstrapping procedures of this kind can be epistemically rational only when some further conditions are met. For example, Glymour proposes the following criterion for determining whether a bootstrapping procedure may be legitimate: `\emph{No Risk, No Gain (NRNG): To test a hypothesis we must do something that could result in presumptive evidence against the hypothesis.}'\cite{Glymour1981-GLYTAE-6} This criterion allows us to see why (A4) is legitimate: there are certainly ways that my memories could have been which would  result in presumptive evidence against the hypothesis (A4), for example if they represented to me a history full of such chaos that it seems impossible that a conscious being could ever have arisen from such conditions.  But this reasoning doesn't work  in the case of (A3), because in that case we are accepting that our memories do \emph{not} represent the actual past, and asserting only that they match the actual past in one very specific way.  So even if my memories look utterly chaotic, (A3) is not disproved, because the actual past could match the relative frequencies in my chaotic memories while being very different from the representation in my memories in many other ways. That is to say, obtaining a contradiction  with (A4) requires us only to determine that \emph{one particular} history makes the existence of a conscious mind unlikely, while obtaining a contradiction with (A3) requires us to determine that \emph{every conceivable} history exhibiting a certain set of relative frequencies  makes the existence of a conscious mind unlikely, and given the enormous size of the possibility space, it's hard to see how we could ever establish the latter with any confidence. So it seems that there is no possible way our memories could have been which would have been incompatible with the hypothesis (A3),  so NRNG does not apply here and thus the boostrapping procedure can't be regarded as providing justification for believing (A3).

\subsection{Transcendental access to the past} 
 
In view of these difficulties, it may seem tempting to say that consulting memories and/or records should not be understood as a measurement whose outcome can be predicted by the theory we are trying to confirm - rather observers  simply have some kind of transcendental access to facts about what has happened in the past. But since we have  agreed that orthodox interpretations do not allow a view from nowhere from which different perspectives can be compared, whatever that transcendental access constitutes, it cannot guarantee that the memories that observers have about events in the past match what previous versions of themselves observed - and thus, in particular, it can't  offer any guarantee that observers will  typically remember relative frequencies which approximately match the probabilities predicted by the theory we are trying to confirm, so without   further details on the nature of this `transcendental' access we simply have no way to say what observers will typically remember. Thus this way of understanding memories and records does not allow us to attach any meaning to the relative frequencies that we happen to remember at any given moment, since the underlying theory plays no role in bringing it about that we remember these relative frequencies rather than others.

Indeed, if this were the way memory really worked in an orthodox interpretation, then what would be the chances that an observer in the world correctly described by that orthodox interpretation would actually come up with a theory   that looks like quantum mechanics on the basis of their memories? Practically zero, one might think - or rather, completely undefined, since the `transcendental access' model is too vague to provide  a probability distribution over what observers will remember -  even the uniform distribution doesn't seem to be justified here. Thus if this is the right way to think about memory in the kind of universe postulated by an orthodox interpretation, it follows that if we believe that we live in such a universe, the fact that our records of observed relative frequencies are consistent with quantum mechanics cannot possibly count as evidence for anything at all.

Note also that in virtue of being \emph{interpretations}, the orthodox interpretations are not just sets of models: they also include   assertions about what those models represent. That is to say, an orthodox interpretation has semantic content, and therefore in order for an agent's beliefs about an orthodox interpretation to be true, presumably the central terms of their beliefs must successfully refer to the   network of  perspectives that the interpretation postulates. But since under the `transcendental access' model an agent   in the type of world postulated by an orthodox interpretation can have the right beliefs about this network of perspectives only by a lucky accident, it's unclear that reference can successfully be established under such circumstances - for as noted by Putnam, it seems plausible to say that a concept fails to refer to an actual entity if the resemblance between the concept and the entity is merely accidental\cite{Putnam1999-PUTBIA}. We will not develop this argument further here, but we note that a very similar objection is  discussed in the context of the Everett interpretation   in ref  \cite{AdlamEverett}.

 \subsection{Other types of evidence}

One way of responding to these concerns would be to observe that there are features of the world other than relative frequencies which count as evidence for quantum mechanics - for example, quantum mechanics is needed to explain the stability of matter\cite{lieb_seiringer_2009}. So maybe we can get empirical confirmation for quantum mechanics  just by observing the stability of matter - after all, this is something that an observer can plausibly observe at a single moment and thus it can be used in empirical confirmation even by agents who do not persist over time. However, this approach also has problems. For a start, it doesn't seem plausible that quantum mechanics is the only possible explanation for the stability of matter - we need evidence that is more specific to quantum mechanics, such as relative frequencies, before we can have good reason to choose it over other possible explanations.  Moreover, the problems we have pointed out for relative frequencies also apply to any other empirical fact that we might hope to use as evidence - for if we can't learn anything about the perspectives of other observers, we can't have any way to know that matter is indeed stable from the point of view of other observers. Perhaps matter is actually not stable relative to most other observers and you just happen to be one of the few lucky observers such that at least momentarily your perception of reality includes a stable macroreality. 
 
 One might perhaps try to argue that it is very unlikely we would find ourselves in an apparently stable macroreality just by chance, and thus argue that we should infer from our observations of a stable macroreality that   macroreality is stable for all observers. But this will not work either: first, because there does not seem to be any way to postulate a probability distribution over different perspectives in such a vaguely defined possibility space, and second, because we can give an anthropic argument. For presumably consciousness can exist for long enough to sustain extended thoughts about the stability of matter only if matter is at least momentarily stable, so given that you are  having thoughts about the stability of matter, then necessarily you are one of the (perhaps very few) observers relative to which matter is stable for at least a moment, so the fact that you are such a observer doesn't licence any inferences about the stability of matter for other observers. Indeed, one can make an orthodox interpretation analogue of the famous Boltzmann brain argument\cite{carroll2017boltzmann}: if there is no knowable observer-independent reality that could ground similarities between different  perspectives, wouldn't it be most rational to suppose that the perceptions of each observer are wildly different and completely unrelated, and thus you are not actually a temporal part of a persisting entity at all,  you are just a fully formed perspective, complete with memories of a past that never really occurred, that happens to exist due to a random fluctuation within the network of separate and unrelated perspectives?  Admittedly it's not very clear how to evaluate the probability of this possibility as compared to the probability that all perspectives are similar in the way postulated by the orthodox interpretations, but that is precisely the problem: if we don't allow any knowable observer-independent facts which are understood to give rise to various different perspectives in some ordered way, there seems no reasonable method to arrive at priors which say anything about the network of perspectives at all. At the very least, the proponents of orthodox interpretations owe us some account of how they have arrived at the set of priors which justifies them in reaching the conclusion that all or most of the perspectives are correctly described by quantum mechanics.

\section{What now for orthodox interpretations? \label{RQM}}

One possible way for proponents of orthodox interpretations to respond to the dilemma that we have posed would be by   walking back the claim that    quantum mechanics is supposed to be a theory of `the whole world' in any sense. That is, they might choose to regard quantum mechanics as simply a convenient codification of the regularities that appear in the facts which are true relative to a single observer: other observers, and other versions of the same observer at other times, will see different regularities and will thus  arrive at different theories which better describe the results that they have seen.  But actually, even this move is not really justifiable. For where did the idea of postulating that everything is relativized to a perspective come from in the first place? It came from treating quantum mechanics as a complete description of the whole world and observing that on this construal, quantum mechanics appears to say things about  the relationships between observations made by different observers which seem difficult to combine into a single `absolute' reality. But if quantum mechanics is only a theory of the world relative to me at this moment, then I am not allowed to attach any significance to the part of quantum mechanics making claims about what other observers will observe: thus  if I believe that all facts are necessarily relative to an observer, then I cannot have any grounds for believing in the features of the theory which are the whole reason for postulating that all facts are relative to an observer in the first place! Thus even retreating to regarding quantum mechanics as a theory which is correct `relative to me now,' will not solve the self-undermining problem. 

So is this the death knell for orthodox interpretations? Well, although we  have criticized orthodox interpretations, we agree that they are based on a correct insight: quantum states are not observer-independent.  Quantum states were postulated in the first place as a way of describing systems relative to external observers, and once we start thinking about how observers themselves should be incorporated into the picture, it becomes very plausible that quantum states should indeed be regarded as relational descriptions in the spirit of Everett's relative state description\cite{Everett}. The orthodox interpretations combine this insight with the assumption that quantum mechanics (or quantum field theory, or quantum gravity) is the correct and final description of reality, and thus arrive at the conclusion that all facts are relative to an observer. Indeed one is usually presented as the direct consequence of the other, without even making the assumption of $\psi$-completeness explicit: for example, Ruyant writes that the relational interpretation proceeds by `\emph{positing that the states quantum mechanics describes are relative to the observer rather than absolute. We should abandon the idea that there is a “view from nowhere.}'\cite{Ruyant2018CanWM} But the proposition that quantum states are relative rather than absolute certainly does not entail that there is \emph{no} view from nowhere: it just entails that a description of of the world `from nowhere' would not contain quantum states as a fundamental entity. There are many ways in which one could have observer-independent facts without having observer-independent quantum states, or indeed states at all. For example, one could have an ontology consisting entirely of pointlike events, as in the Bell flash interpretation\cite{Tumulka_2006,tumulka2020relativistic}.

Why, then, do proponents of the orthodox interpretations appear to ignore all of these intermediate possibilities? In large part it is because they are aiming to interpret the existing mathematical framework of unitary quantum mechanics without adding anything new, and since the existing mathematical framework in its most common formulation describes reality in terms of states, it seems to follow that if they wish to eliminate observer-independent facts about states, they will necessarily have to eliminate all observer-independent facts whatsoever. But this motivation is flawed for several reasons. For a start, there are other formulations of  quantum mechanics which don't describe reality in terms of states - for example, the path integral formulation\cite{feynman2010quantum}. That formulation potentially offers a route to interpreting the existing mathematical formulation in a way which does not postulate observer-independent states, but does nonetheless postulate knowable observer-independent \emph{facts}, i.e. facts about the path integral. Second, the drive to maintain quantum mechanics in its standard formulation often seems to be driven largely by an aesthetic preference for the pleasingly simple unitary evolution, but we should not let aesthetic preferences blind us to all other considerations, such as the serious epistemic problems discussed in this article. Furthermore, there are ways to add something to the mathematical framework of unitary quantum mechanics without losing the aesthetically pleasing unitary evolution: for example, Kent's solution to the Lorentzian classical reality problem\cite{Kent, 2015KentL, 2017Kent}  involves allowing the wavefunction to evolve unitarily until the end of time and then performing a final measurement which selects and actualises a definite course of history, thus providing `observer-independent facts' without compromising any of the mathematical structure of unitary quantum mechanics.

Another major motivation for proponents of orthodox interpretations is the desire to avoid certain features which are regarded as undesirable - such as nonlocality, contextuality, retrocausality and so on. But this way of thinking gets the priorities wrong.  We can do without locality if we have to, but we certainly cannot do without the possibility of empirical confirmation. Nonlocality may present  technical challenges, but it does not pose this kind of existential threat to the status of the theory as empirically confirmed.\footnote{We also note that in some cases it is argued that we need locality because otherwise the theory will be inconsistent with relativity, and therefore we have to accept  an orthodox interpretation if we want to have any hope of a successful unification with gravity. But this is a misconception arising from the fact that early nonlocal interpretations, like the GRW collapse model\cite{Frigg2009} and the de Broglie-Bohm approach\cite{holland1995quantum}, required an instantaneous wavefunction collapse or update on a spacelike hyperplane. Later nonlocal interpretations, like the relatistic Bell flash model\cite{Tumulka_2006,tumulka2020relativistic} and Kent's Lorentzian quantum reality model\cite{Kent, 2015KentL, 2017Kent} do not require a preferred reference frame and are explicitly Lorentz-covariant, thereby demonstrating that consistency with relativity does not force us to adopt an orthodox interpretation. } So rationality demands that we prioritise securing the evidential basis of the theory over trying to make it local -  or non-contextual, or non-retrocausal etc. Indeed, given that epistemic rationality is a theoretic virtue which science must surely prioritise above all else, it is hard to see how interpretations which undermine the evidential basis for the theory could ever be preferred as long as there are any other viable options.

\subsection{Altering orthodox interpretations}

So what options do we have if we want to come up with a version of an orthodox interpretation which overcomes the problems set out in this article?  An obvious first step is to allow that there exist some mind-independent absolute facts, albeit not facts about the physical states of systems. This seems to be what Ruyant has in mind when he writes that  `\emph{(RQM's) general ontological claims (“what the world is like” according to RQM) should be understood as being absolutely true in order to avoid the self-refutation problems associated with relativism, and the relativist stance should be reserved to statements about particular entities or facts in the world.}''\cite{Ruyant2018CanWM} This certainly seems like a step in the right direction - at least that way there is something which could possibly be the subject of the theory we are aiming to confirm!  However,  simply saying that the `general ontological claims' of an orthodox interpretation are an observer-independent fact is not adequate, because typically the ontology of an orthodox interpretation consists simply of observers plus perspectives relative to them, but we have seen in this article that in order for these approaches to allow that quantum mechanics is   subject to empirical confirmation there must be something more - some kind of unifying structure which underwrites intersubjective agreement in at least certain cases. 

Now,  one of the motivations for the denial of observer-independent facts in orthodox interpretations is the fact  that any specific proposal for an observer-independent reality will most likely need to have properties like nonlocality which many people find unappealing, so one might naturally hope to take the minimal route here by keeping one's preferred orthodox interpretation as it is and simply adding the hypothesis that there exists some kind of unifying structure which guarantees intersubjective agreement in most cases, without saying anything more about the nature of this unifying structure. For example, QBists might perhaps try to argue that the unifying structure in question is their ineffable  `external reality' - there exists something which brings about intersubjective agreement but we are not allowed to ask how or why. But this is not good enough.  For a start, in order to verify that this `unifying structure' can actually play the desired role in making empirical confirmation possible, we need to know the limits of its promise of intersubjective agreement: presumably this structure still cannot guarantee intersubjective agreement in special cases like the Frauchiger-Renner experiment, so what are the limits of the intersubjective agreement and how reliable is it? Moreover, once we've added to our preferred orthodox interpretation the stipulation that different  observers can be expected to agree  most  of the time, one might wonder why exactly we need to keep insisting that everything is relative to an observer: it would seem simplest to just accept that whatever it is they agree on is really the fundamental stuff of reality. Thus in order to find a route which both achieves enough intersubjective agreement to make empirical confirmation possible but also maintains  the  relevance and usefulness of an observer-dependent approach, it would be necessary to give a reasonably detailed account of the specific circumstances in which perspectives do and do not agree. Something like the QBist's ineffable `external reality' is not sufficient to solve the problem.

 Rovelli has recently suggested a new postulate for RQM aimed at solving the problem we have set out in this article: 
 
\begin{definition} 

\textbf{Cross-perspective links (CPL):} In a scenario where some observer Alice measures a variable V of a system S, then provided that Alice does not undergo any interactions which destroy the information about V stored in Alice's physical variables, if Bob subsequently measures the physical variable representing Alice's information about the variable V, then Bob's measurement result will match Alice's measurement result

\end{definition}

  That is to say, when Alice sees a definite value in a measurement, that piece of knowledge is encoded in her physical state in a way that is accessible to Bob. We agree with Rovelli that this is exactly what is needed in RQM to solve the problems that we have raised in this article, and  a companion article\cite{pittphilsci20379} discusses the consequences of implementing this postulate. In particular, it is important to note that adding this postulate requires us to make some changes to existing accounts of the ontology of RQM: it is no longer the case that quantum mechanics does not describe an observer-independent reality, because it is now the case that  the values of the variables of a system ($S$) which become definite during an interaction with another system  (Alice)  are   observer-independent at least in the sense that that any other observer can in principle find out these values by an appropriate measurement on either $S$ or Alice. So although RQM can potentially solve the problem we have set out in this article, it can only do so by ceasing to be what we have described as an `orthodox interpretation.' 

Can other orthodox interpretations adopt something like \textbf{CPL}? This seems unclear. The postulate works in RQM because Rovelli and co-authors have always maintained that all physical systems count as possible `observers,' and therefore it is possible to to use interactions between observers and systems in RQM to define an objective reality consisting of a set of mind-independent events.  On the other hand, approaches like QBism and neo-Copenhagen views  are more inclined toward the view that only conscious minds count as observers, and therefore we don't have the option of  using interactions between observers and systems  to define a mind-independent set of events, since every such interaction necessarily involves a conscious observer. Moreover, RQM has always been associated with the naturalistic principle that `information is physical' so it is actually very much in line with the underlying motivations of the approach to insist that information possessed by observers should always be accessible to other observers by appropriate physical interactions. Conversely,  neo-Copenhagen approaches and similar views are often motivated by the idea that information itself is fundamental and the physical world is in some sense emergent from it, and therefore there is no  physical world available to ground the expectation of agreement between perspectives, so it seems much harder in such cases to come up with any plausible justification for something like \textbf{CPL}. 

At any rate, even if these orthodox approaches are willing to adopt something like \textbf{CPL} this will definitely involve some quite significant changes in the metaphysics and ontology associated with these views. Proponents of orthodox interpretations cannot simply  postulate this coordination between perspectives ad hoc without saying anything about the source of it - explaining the source of regularities is, after all, at the heart of the scientific endeavour according to most scientific realists!  And it is particularly crucial in this case because  the intuitively obvious explanation for the coordination would be to say that there exists an observer-independent underlying state of S and thus the measurement results obtained separately by Alice and  Bob agree precisely because they are both probing the same underlying state. Indeed, in the absence of any other concrete explanation for the coordination, inference to the best explanation would seem to compel us to draw this conclusion. But the nonexistence and/or inaccessibility of observer-independent  reality is a central principle of most of these orthodox interpretations - a version of QBism or the neo-Copenhagen approach which allowed observers to probe a mind-independent underlying state would be so far removed from the original that it could hardly be said to be the same thing at all - so if proponents of orthodox interpretations are to accept something like  \textbf{CPL}, it will be vital to come up with some equally compelling explanation for the coordination in order to block this inference to the best explanation.

There would seem to be roughly three ways to provide such an explanation, each corresponding to dropping or adjusting one of the three  fundamental characteristics of an orthodox interpretation set out in section \ref{orthodox}. First, we could drop the assumption of unique measurement outcomes, which would   lead us to the Everett interpretation or similar. Second, we could drop the requirement that quantum mechanics is universal  and argue that at least some parts of macroscopic reality cannot be described quantum-mechanically, since they exist in a single definite way for all observers. However, we would then have to specify where the line between the quantum-mechanical and the non-quantum-mechanical parts of reality falls, which is just the standard measurement problem, and therefore we would presumably be forced to use some other, non-orthodox interpretation of quantum mechanics to draw this line. Third, we could drop the requirement that quantum mechanics does not describe an observer-independent external reality and add in some kind of observer-independent structure which coordinates perspectives in the required way. However,  once we move to this kind of picture  we have a compelling alternative ontology available to us: rather than insisting that everything is relative to a perspective, we find ourselves with an ontology composed of events which are observer-independent in the sense that any other observer can in principle find out about the values of the variables realised in the events, together with some kind of coordinating structure linking these events together. And once we move to that kind of picture it becomes very natural to wonder if we could perhaps find a way of describing the distribution of events all-at-once using the coordinating structure, rather than using partial, relational descriptions as in quantum mechanics - which is to say, it becomes less clear that we can still maintain that standard quantum mechanics is complete and universal. So this way of thinking leads us inevitably away from the kind of picture associated with the orthodox interpretations and towards some other kind of realism which deserves to be fully explored rather than hidden under the veneer of an orthodox interpretation.

\section{Conclusion} 

 It is understandable that proponents of the  orthodox interpretations wish to insist that there are no knowable observer-independent facts: for if they allow knowable observer-independent facts then they will have to say something about what these observer-independent facts are, and   various quantum no-go theorems make it clear that it will be difficult to do so without postulating something in the realm of  hidden variables, nonlocality, retrocausality, or superdeterminism - all things that proponents of orthodox interpretations typically wish to avoid at all costs. But  in order for it to be rational for us to believe an orthodox interpretation, something must be done about empirical confirmation, and therefore  these approaches must be supplemented with some mechanism for  selecting and actualising measurement outcomes in such a way as to provide at least some minimal level of intersubjective agreement across different relative descriptions.  Moreover the observer-independent structure responsible for the coordination cannot simply be  `the wavefunction' since then the view would collapse into the Everett interpretation, and therefore it would seem that something must be added to the existing formalism if we are to make sense of empirical confirmation within these approaches.

Now, of course we do not mean to suggest that the formalism is unusable in its current form: in most circumstances (i.e. outside of special cases like the Frauchiger-Renner experiment) we know perfectly well how to use quantum mechanics to make unambiguous predictions. So it is reasonable for a physicist focused on applications to simply have no interest in the nature of the underlying observer-independent reality, and we have no objection to that pragmatic choice. However, we do object to the form of argument which involves first proposing an orthodox interpretation and then using that orthodox interpretation to dismiss all attempts to describe an observer-independent underlying reality. And we particularly object to the claim that  such attempts are driven by attachment to   na\"{i}ve classical ideas - the commitment to an observer-independent reality is not simply a pre-scientific prejudice, rather it is a reasoned position based on the observation that some kind of coordination between perspectives is necessary to ground the rationality of science. Therefore there are good scientific reasons for being committed to the existence of an observer-independent reality (and moreoever, one that is not ineffable), and thus also good scientific reasons for trying to understand the nature of that observer-independent reality. 
 
We also do not mean to suggest that because orthodox interpretations make it impossible for us to obtain confirmation for scientific theories like quantum mechanics, the world can't  be the way they describe. Rather we claim that it can't be rational to \emph{believe} that the world is the way they describe \emph{as an interpretation of quantum mechanics}. Of course we accept that it could be the case that really there are no observer-independent facts - indeed, there might even be good reasons independent of physics to believe that to be the case. Instrumentalism, antirealism, idealism, phenomenalism and the like have certainly had many defenders over the centuries, and the concerns raised in this article don't threaten the traditional arguments for them, because people do not usually argue for these positions on the basis of a specific scientific theory.  We simply contend that it is not rational to believe that the world is the way the orthodox interpretations describe \emph{because of quantum mechanics}, since it is not rational to hold a belief which undermines its own evidentiary basis. In order to do science we have to assume that there are appropriate preconditions in place for us to successfully learn about reality by means of empirical observations, and therefore we should insist that any interpretation of a theory we thereby arrive at should at least approximately uphold those preconditions and explain how it is possible for us to   gain evidence for the theory in question.

 We finish by recalling that, as argued in ref \cite{AdlamEverett}, that a very similar problem of confirmation  may apply to the Everett interpretation, and  in fact we would suggest that   problems connected to confirmation are likely to arise quite generally in all interpretations of quantum mechanics which deny the common-sense idea that macroscopic measurement events have unique outcomes which are the same for all observers. For after all the theory of quantum mechanics was arrived at by analysing sets of measurement outcomes  under the assumption that these outcomes were unique and the same for all observers;  orthodox interpretations and the Everett interpretation  both take for granted that  the empirical confirmation that has been accrued for quantum mechanics under this assumption can simply be transferred to a completely different context where the assumption no longer holds, but as we have seen in detail in this article, empirical confirmation will not always be maintained under a significant reinterpretation of the theory, and therefore the proponents of these interpretations should be obliged to provide an account of how empirical confirmation is supposed to work in the new setting that they propose.  

Indeed, we would suggest that  showing explicitly how the interpretation allows the theory to be empirically confirmed should really be compulsory for anyone who wants to propose an interpretation of any scientific theory - an interpretation should not be drawn entirely from the mathematics of a theory, but should also demonstrate a clear link to the empirical data which is our evidence for that theory.  Paying inadequate attention to the problem of empirical confirmation results in the formulation of  interpretations that are interpretations of the \emph{mathematics} but which fail to be compelling interpretations for the \emph{empirical evidence} which is the reason for the mathematics in the first place: we would like to see discussions of the interpretation of quantum mechanics place more emphasis on the empirical evidence for the theory, and we would even venture to hope that centering the evidence in this way might provide a way out of the long-standing impasse over the interpretation of quantum mechanics.

\section{Acknowledgements}

Thanks to Carlo Rovelli for very helpful discussions. This publication was made possible through the support of the ID 61466 grant from the John Templeton Foundation, as part of the “The Quantum Information Structure of Spacetime (QISS)” Project (qiss.fr). The opinions expressed in this publication are those of the author and do not necessarily reflect the views of the John Templeton Foundation.

 \bibliographystyle{unsrtdin}
 \bibliography{newlibrary12}{}

\end{document}